\documentclass{olplainarticle}

\usepackage{tikz}
\usepackage{hyperref}
\usepackage{url}
\usepackage{mdframed}
\usepackage{pgfplots}

\pgfplotsset{compat=1.14}

\newmdenv[linecolor=gray,backgroundcolor=gray!20]{function}
\newmdenv[linecolor=teal,backgroundcolor=teal!20]{problem}

\newcommand{\polis}{\href{https://pol.is/home}{{\tt Polis}}}
\newcommand{\polaire}{\href{https://polaire.quebec/}{{\tt Polaire}}}
\newcommand{\allourideas}{\href{https://pol.is/home}{{\tt AllOurIdeas}}}
\newcommand{\liquidfeedback}{\href{https://liquidfeedback.com/en/}{{\tt LiquidFeedback}}}
\newcommand{\remesh}{\href{https://www.remesh.ai/}{{\tt Remesh}}}
\newcommand{\kieskompas}{\href{https://www.kieskompas.nl/en}{{\tt Kieskompas}}}
\newcommand{\habermas}{\href{https://github.com/google-deepmind/habermas_machine}{{\tt Habermas Machine}}}

\DeclareMathOperator*{\argmax}{arg\,max}
\DeclareMathOperator*{\argmin}{arg\,min}

\title{
Computational
Challenges \\
in Scaling
Democratic Deliberation\footnote{Version 4, May 2026.}
}

\author{\href{www.davidegrossi.me}{Davide Grossi}}
\affil{
Bernoulli Institute for Mathematics, Computer Science, and Artificial Intelligence \\
{\em University of Groningen} \\
Institute for Logic, Languange and Computation \\
Amsterdam Center for Law and Economics \\
{\em University of Amsterdam}
}


\begin{abstract}
The paper provides an overview of core functionalities that digital democracy software needs to provide in order to support democratic deliberative processes at scale. Developing these functionalities poses novel computational challenges and requires algorithmic solutions to interesting mathematical problems. The aim of the paper is to break the first ground towards a structured inventory of such problems, and to position possible approaches to them within current academic research in computer science and artificial intelligence.  
\end{abstract}

\begin{document}

\flushbottom
\maketitle
\thispagestyle{empty}




\begin{flushright}
\hfill\parbox{210pt}
{
{
{\em The benefits from discussion lie in the fact that even representative legislators are limited in knowledge and the ability to reason.
No one of them knows everything the others know, or can make all the same inferences that they can draw in concert. Discussion is a way of combining information. 
}
}

\citet{rawls2017theory} \\
}
\end{flushright}

\begin{flushright}
\hfill\parbox{210pt}
{
{\em The discipline of computing is the systematic study of algorithmic processes
that describe and transform information.
}

\citet{denning1989computing} \\
}
\end{flushright}

\section{Introduction}

\paragraph{Context: democratic responsiveness.}
Democratic policies are policies that track and respond to the preferences and views of citizens, and do so in a way that treat citizens as peers---that is, responding to democratic principles of fairness, equitability, inclusivity, and the like. To say it with political scientist \citet{dahl2008polyarchy}, their key feature should be: 
\begin{quote}
   ``the continuing responsiveness [\ldots] to the preferences of its citizens, considered as political equals''.  
\end{quote}
The aim of digital deliberation platforms (in short, DPs) is to leverage digital technology to support deliberative processes at scale. In doing so, they provide a novel path towards democratic responsiveness by enabling means to elicit and aggregate the views/opinions/positions/preferences of citizens on issues that they consider important to the lives of their communities. Deliberation is the go-to process that provides this type of rich responsiveness in small groups, 
but the only way to realistically achieve a comparable level of responsiveness at scale requires digital media and algorithmic support. It has also become evident that current technology for mass interaction---such as social media---cannot serve that purpose, because its design is driven to the largest extent by commercial aims \citep{narayanan2023understanding}. 

\paragraph{Problem: legitimacy of DPs.} 
Online platforms for citizens' participation in policy-making have been object of substantial multi-disciplinary research in the last decade. Such research has typically focused on: spelling out the potential of such technology to improve democratic legitimacy; identifying normative criteria that the technology should adhere to in order to be effective as well as conditions for its success; the analysis of case studies of real-world deployment of the technology. Importantly, however, virtually no research (with very few exceptions) has been conducted on what algorithms should actually be used to power DPs. Surprisingly, even government-commissioned scientific reviews (e.g., \cite{rosa2025harnessing}) that explicitly aim at generating momentum to improve the prospects of the technology for citizens hoover above this algorithm-design issue. 
Such an omission is striking because DPs are run by algorithms and the quality of any DP will, therefore, inevitably depend on the quality of the algorithms that run it `under the hood'.

So, why is algorithmic research on DP essential for their democratic legitimacy? Let me make a simple example. There is wide consensus that any DP should guarantee inclusivity for all participants. The question then arises of: {\em What algorithms best guarantee participants' inclusion?}. This, in turn, calls for answers to these related questions: {\em What is inclusion in the context of DPs?} and {\em How, or to what extent, can such notion be mathematized so that algorithms can be designed to achieve it?}. Answering these questions is not simple, and answers will not be unique. But making them explicit is the only way to build confidence in a technology with large social-impact potential such as DP and, ultimately, guarantee the transparency of its design.

\paragraph{Focus: computational problems in DPs.}
A computational problem is a task that is amenable to mathematical formulation and to being solved using an algorithm. DPs need to be able to solve many such problems to deliver deliberation-support functionalities. To make a simple example, suppose we are running an online deliberation and consider a question of this type: {\em given the ideas that have so far been proposed by participants during the deliberation, how can we select a sample of them which is representative of the views of all users?} In other words, we are given an input (the current proposals made by participants and their attitudes towards them) and we are asked to find an output (a sample/subset of all the proposals) which satisfies a specific objective (a form of representativity). By calling this problem {\em computational}, I emphasize that we are looking for an algorithm to solve it, that is, a step-by-step procedure. The quest to operationalize a democratic value such as representativity in a DP gives rise to a number of computational problems like the above ones. The aim of this paper is to break the first ground towards an initial inventory of such problems based on existing approaches, which can hopefully stimulate further research on the algorithmic layer of DPs. As such, the paper zooms in on deliberation support as part of a broader research agenda on digital democratic innovations, that was outlined by \citet{grossi2024enabling}. 

The computational problems I discuss will be formulated in
\colorbox{teal!20}{teal-colored boxes} for easy identification.  The formulations I propose for those problems will necessarily be high-level. Translating them to mathematically defined problems will require extra research and is beyond the scope of this paper. The formulations I propose aim at putting such problems on the map in a systematic fashion, and stimulate precisely the sort of research needed to define them, and subsequently find algorithmic approaches to them.

\paragraph{Scope.} Finally, it is worth stressing that I do not claim exhaustiveness with respect to the challenges I identify. Also, I focus specifically on DPs that are designed for asynchronous and text-based interaction, which are the only ones that can, arguably, scale to large groups (tens of thousands users). The paper is therefore not concerned with platforms such as, for instance, the Stanford Deliberation Platform\footnote{\url{https://stanforddeliberate.org/}}, which is meant for interaction by 8-15 people. Also, while some DPs support decision-making on top of deliberation (e.g., via voting like \liquidfeedback), others do not and aim just at eliciting rich information from participants (like \polis, \allourideas, or \polaire). In this paper I will not touch upon decision-making functionalities and focus exclusively on those functionalities of DPs that are designed to support the deliberative element of the platform.


\section{Algorithmic governance of large-scale democratic deliberation}

The form of deliberation supported by DPs is, in its nature, necessarily different from face-to-face verbal deliberation: the deliberative process occurs over a long stretch of time, a potentially large crowd takes part in it, and large amount of information is produced in the process, of which participants need to make sense. Such a process, in order to be fruitful and to respond to democratic criteria, needs to be governed. Manual bureaucratic governance of such large-scale deliberative processes is simply infeasible. Algorithms need to be used to govern them.

\subsection{The {\em deliberation-support loop}}

In this paper I put emphasis on DPs where the need for manual moderation is kept as limited as possible. In particular, I focus on what I believe to be the core governance aspect in deliberation: `who says what when'. In DPs this is taken care of collectively by designing and structuring the conversation trough what I refer to as {\em deliberation-support loop} and the algorithms that power it---what \citet{liquid_feedback} refer to as {\em collective moderation}. It is suggestive to think of the deliberation-support loop, with its underlying algorithms, to be for online deliberation what rules of order\footnote{For instance, Robert's rules of order \citep{robert2020robert}.} are to physical deliberative assemblies like parliaments.

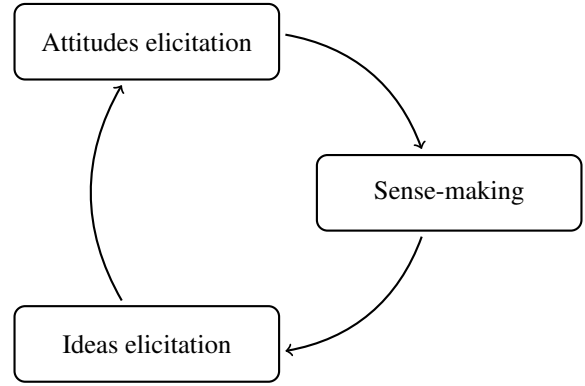
\begin{figure}
\centering

\begin{tikzpicture}[
  box/.style={
    rectangle, draw, thick, rounded corners,
    minimum width=3.5cm, minimum height=1cm,
    align=center
  },
  ->, thick, shorten >=2pt, shorten <=2pt
]
\node[box] at (0,0) (ideas) {Ideas elicitation};
\node[box] at (0,4) (attitudes) {Attitudes elicitation};
\node[box] at (4,2) (sensemaking) {Sense-making};

\draw[bend left=30] (ideas) to (attitudes);
\draw[bend left=30] (attitudes) to (sensemaking);
\draw[bend left=30] (sensemaking) to (ideas);

\end{tikzpicture}
    \caption{The deliberation-support loop in online deliberation platforms.}
    \label{fig:loop}
\end{figure}

\medskip
To govern a large online deliberative process there are, in my view, three core functionalities that a deliberation-support loop provides:
\begin{enumerate}
    \item \fbox{Ideas elicitation}. With `idea' I mean any text (free or possibly somewhat structured according to specific pre-determined formats), which can represent a suggestion, a proposal, or a statement that a user may want to contribute to the collective deliberation. So, I use the term `idea' here as a catch-all term for any text-based contribution that participants may input in a DP. A key feature of DPs is that the set of ideas to be deliberated upon is not fixed and, although an initial set of them may be seeded by the initiators of the deliberation, an explicit goal of the process is to have participants contribute their own ideas towards the deliberation.
    \item \fbox{Attitudes elicitation}. With `attitude' I mean an evaluative opinion about proposed ideas. The format of such attitudes can vary from approving or disapproving proposed ideas (like in \polis, or \liquidfeedback), to ranking them (like in \allourideas), to possibly assigning scores to them. Such attitudes are collected in order to obtain feedback on how the deliberating group feels about the ideas proposed.
    \item \fbox{Sense-making}. With `sense-making' I mean any data analysis performed on collected attitudes towards proposed ideas, which aims at displaying (on a normal screen) a unified, easy to comprehend, `common-ground' view of the current state of the deliberation. In \polis~this takes the form of, among others, a so-called `opinion landscape' \citep{small2021polis}. In \liquidfeedback~and \allourideas~it takes the form of special types of rankings \citep{liquid_feedback,salganik2015wiki}.
\end{enumerate}
We should think of these functionalities as operating iteratively in phases through a loop (see Figure \ref{fig:loop}): ideas are elicited from the group; then attitudes towards those ideas are elicited; finally the resulting information is presented back to the group in some aggregated format to feed the deliberation further; in response to such an aggregation, new ideas are elicited; and the loop continues for a pre-determined amount of time. 

\medskip

In the remaining of the paper I will unpack each of the above-mentioned functionalities and provide an inventory of mathematical and computational problems that underpin the algorithmic delivery of such functionalities, and do so in ways that can be considered appropriate from a democratic standpoint.

\paragraph{Mathematical notation.}
At this point it is useful to introduce a minimal amount of mathematical notation. To refer to an arbitrary proposed idea I will use $p$ and write $P$ for the set of all ideas that have been proposed. To refer to an arbitrary participant I will use $i$ and write $N$ for the set of all participants. Attitudes can take a variety of forms, for example: approval/disapproval (`thumbs up/thumbs down'); rankings (`$p_1$ is better than $p_2$'); judgments (`$p$ is decent', `$p_2$ is excellent'). To illustrate what discussed in the paper I will resort to the simplest possible type of attitudes that a participant can entertain towards an idea: explicit approval. So, participant $i$'s attitude towards a proposed idea $p$ is denoted $A_{ip}$ and takes the value of $1$ (if $i$ approves of $p$) and otherwise $0$ ($i$ does not approve of $p$), and possibly value `undefined', which I denote $\star$ ($i$'s attitude towards $p$ is unknown).\footnote{Clearly, assignments on different scales are possible, but for the purposes of this paper, the set $\{1, 0, \star \}$ suffices.} I will refer to a matrix that contains $\star$ entries as {\em partially elicited}. I will also make use of standard set-theoretic notation throughout the paper. However, the mathematics in the paper could be disregarded while still getting its main messages.

\subsection{Ideas elicitation}

A first design choice involved in the development of a DP concerns the kind of contributions---what we called `ideas'---the platform aims at collecting from participants. This concerns, essentially, the question about {\em the type(s) of input the platform accepts from participants}.
Ideas are here considered to be just pieces of text that participants may input in the system. These inputs, however, can be more or less structured and can be more or less interrelated. Several formats can be observed in existing platforms:
\begin{description}
    \item[Independent inputs.] All ideas are independent pieces of texts entered in the system. This is the case, for instance, in \polis.
    \item[Interdependent inputs.] Ideas may exhibit explicit interdependencies where an idea explicitly refers to another one, for instance, in an attempt to improve it. This type of explicitly interdependent input is possible, for instance, in \liquidfeedback.
    \item[Structured inputs.] Ideas need to follow a specific format, for instance an argumentative one, like following a premise-conclusion scheme. This type of input is on the one hand constraining (as it may straitjacket the way in which participants input their ideas), but may on the other hand facilitate the analysis of contributed ideas (see Section \ref{sec:sensemaking}).
\end{description}
In general, whatever structure may ideas be subjected to, may facilitate their analysis while, on the other hand, somewhat constrain their form.  


\subsection{Attitudes elicitation} \label{sec:attitudes}

As ideas are elicited, DPs need also to elicit what participants think of those proposed ideas. It is useful to think of the output of this phase as a matrix where columns are ideas and rows are participants. A cell $A_{ip}$ in this {\em attitudes matrix} ${\bf A}$ collects numerical information about what $i$ thinks about $p$: for instance, whether $i$ approves or disapproves of $p$. See figure \ref{fig:matrix}.

\begin{figure}[t]
\begin{align*}
\hspace{1cm}
\begin{pmatrix}
A_{11} & A_{12} & \ldots & A_{1m}\\
A_{21} & A_{12} &  \ldots & A_{2m} \\
\vdots & \vdots & \ddots & \vdots \\
A_{n1} & A_{n2} &  \ldots & A_{nm}
\end{pmatrix}
\hspace{2cm}
\begin{pmatrix}
\star & 1 & \ldots & \star\\
0 & 0 &  \ldots & \star \\
\vdots & \vdots & \ddots & \vdots \\
1 & \star &  \ldots & 0
\end{pmatrix}
\end{align*}
\caption{Attitudes matrix (left) for $m$ ideas and $n$ participants. In the simple approval/disapproval case (right), each entry $A_{ip}$ (with $1 \leq i \leq n$ and $1 \leq p \leq m$) can get value $1$ (approval), $0$ (disapproval), or $\star$ (undefined).}
\label{fig:matrix}
\end{figure}

The attitudes elicitation phase of the deliberation-support loop can be viewed as a form of questionnaire, or survey, in which participants are asked what they think of ideas proposed by other participants. Crucially, however, it is a form of survey in which the number of items to be submitted to respondents may be too large for respondents to realistically respond to all items: in a large online deliberation process, the number of ideas proposed by participants can easily exceed what participants can possibly consider when asked what they think of those ideas. As we cannot query all participants on all ideas, the natural question arises of {\em how should participants' attitudes be collected}.
Answering this question means, essentially, to come up with an approach to manage participants' attentions towards proposed ideas. There are two main approaches to answer this question.
\begin{description}
    \item[Global querying.] This approach essentially assumes that participants will express their attitudes based on what it is displayed on the screen, and ideas are displayed to all participants in one same way. This boils down to allocating positions on the screen to the proposed ideas, typically in the form of a ranking (akin to the way in which results are displayed by a search engine after a query on the web). This is one of the approaches used in, for instance, \liquidfeedback (see also Section \ref{sec:sensemaking} below). Many considerations are relevant here. First, the ranking should provide a `faithful' rendering of the attitudes elicited so far (I will come back to this specific issue again in Section \ref{sec:sensemaking}), but also make sure the ranking supports further elicitation of ideas and attitudes by participants, so that the deliberation can develop further in the most promising directions. These objectives could be formulated in many different ways, e.g.: guarantee that ideas that obtained so far relatively low exposure can appear high in the ranking. Even the formalization of this type of criteria is an open research question. 
    In any case, global querying approaches can be thought of responding to the following computational problem:
    \begin{problem}
    {\bf {\em Rankings for elicitation-support}}:
    {\em Given} a partially elicited attitude matrix, {\em find} a ranking of the set of all ideas that `best supports' further elicitation.
    \end{problem}
    
    \item[Local querying.] This approach refrains from finding one unique way to display ideas to participants, and rather submits different ideas to different participants for further attitude elicitation. This can be achieved in many ways with various levels of adaptivity, and typically using different forms of randomization \cite{halpern2023representation,lindeboom2025voice}.\footnote{It is worth observing that a global approach to querying, for instance through ranking-generation, can be used to implement a local querying strategy where ideas are submitted to participants with probability proportional to their position in the ranking. This is actually how the local querying approach is implemented in \polis~\cite{small2021polis}.} This is the approach used, for instance, in \allourideas~ and \polis, where the problem is referred to as {\em opinion routing}: 
    \begin{problem}
    {\bf {\em Adaptive ideas elicitation}}:
    {\em Given} a partially elicited attitude matrix, {\em select} a set of ideas about which a specific set of participants should be queried next to `best support' the deliberative process. 
    \end{problem} 
    This a task that falls squarely within the remit recommendation algorithms \cite{narayanan2023understanding}. Essentially it concerns a form  of adaptive recommendation based, however, on the optimization of objectives conducive to democratic deliberation. 
    
\end{description}


\subsection{Sense making} \label{sec:sensemaking}

A crucial functionality of large-scale deliberation support is arguably sense making. In large-scale deliberation, participants are necessarily exposed only to {\em local} views of the state of the deliberation: essentially the ideas that they have contributed and have directly been responding to. Sense making concerns the ability of the platform to provide participants with a {\em global} view of the state of the deliberation, which can be functional to a healthy development of the deliberative process.\footnote{It is worth noting that this is a type of functionality conspicuously missing in mainstream social media.} The sense-making functionality can therefore be thought of responding to the following question: {\em How should the current state of the deliberation be fed back to participants?}.

Clearly, such representation should be a function of the ideas and attitudes elicited at any given time in the process. Crucially, however, such analysis should respond to democratic requirements: {\em Are all ideas and users `fairly' treated in such representations? Is the mode of representation inclusive towards all ideas and users?} So, the sense-making functionality can be thought of as a form of real-time data analysis that, however, needs to respond to democratic desiderata because it provides the basis for further deliberation. The feedback that this type of analysis provides to participants is crucial for them to make sense of the existing contributions and to stimulate a constructive development of the deliberation towards further ideas. It is what provides the `common-ground' that any DP needs to construct for its users.\footnote{
A recent technical report by the Joint Research Center of the European Commission \citep{rosa2025harnessing}, reviewing the use of digital deliberation technology in the context of the Conference on the Future Europe, explicitly highlighted the importance of adequate analytical tools as one of the six key takeaways:
\begin{quote}
    Effective and robust analysis systems are key in online citizen engagement processes to infer the most salient contributions and/or clusters of contributions; to provide valuable insights into citizens' concerns and preferences and to help policymakers identify agreement and disagreement on complex issues.
\end{quote}
}
Achieving this, we will see, requires the solution of interesting computational problems.



\section{Mathematics of sense making in the deliberation-support loop}

Besides providing elementary statistics about elicited attitudes towards proposed ideas (e.g., what are the ideas that elicited largest/smallest support, or that appear to be most controversial), the goal of sense making in the context of democratic deliberation is arguably to provide a {\em fair representation} of the state of the deliberation, in pretty much the same way in which we think of parliaments as needing to be representative of the whole population. Crucially, such {\em fair representation} of the state of the deliberation needs to be computed under {\em incomplete information}, and requires therefore the need of some form of estimation of either individual data points or global metrics. Handling this type of data incompleteness (or sparsity) is a fundamental challenge and I will articulate it in more detail in Section \ref{sec:inc}.



\medskip

One can identify four main types of representations used for the purpose of sense making. These are not necessarily exclusive and may be interfaced.
\begin{description}
    \item[Representative sets of ideas.] A subset of ideas (of a given size) is selected among all ideas proposed in a way that is `representative' of the state of the deliberation.
    \item[Representative rankings of ideas.] Proposed ideas are ranked in a `representative' way as to capture salient features of the patterns of support for ideas (e.g., in \liquidfeedback).
    \item[Idea landscapes.]  Ideas and/or users in a space and cluster them via some notion of proximity (e.g., in \polis). 
    \item[Rich representations.] Linguistic or argumentative patterns that emerge from contributed ideas are distilled and summarized via representations that process the textual information contained in ideas, as well as the elicited attitudes towards those ideas. Unlike the three previous approaches, this one necessarily requires handling also linguistic information contained in contributed ideas, besides support information. Such processing can involve state-of-the art natural language processing techniques, such as in the \habermas. 
\end{description}
In the remaining of this section I expand on each of these approaches.

\subsection{Representative sets of ideas}  \label{sec:rep_set}

To the best of my knowledge, this approach has been investigated almost exclusively in academic work where researchers have studied the following computational problem:

\begin{problem}
    {\bf {\em Representative slates}}:
    {\em Given} a partially elicited attitude matrix, {\em find} a subset---or {\em slate}---of proposed ideas of a predetermined size $k$ that `fairly represents' all participants attitudes.
\end{problem}
The size $k$ can be viewed as a parameter dictated, for instance, simply by how many ideas can be displayed at a same time in a comfortable (to the viewer) way. So, the goal here is to find $k$ ideas among all the proposed ones that can be considered to be particularly representative of how participants think about what has been deliberated so far. 

Approaches to the above question typically rest on assigning a suitably defined `representativity' score $S$ to sets of ideas, and then choosing the $k$ ideas that maximize such score. Mathematically, the above question is answered by choosing sets of representative ideas that are (possibly approximate) solutions to the following optimization problem:
\begin{equation}
    \argmax_{X \subseteq P, |X|=k} S(X),
\end{equation}
where $S$ is a {\em scoring} function that maps sets of ideas to numerical (typically, rational) values.\footnote{These are known as {\em Thiele} methods in the literature on committee voting \cite{lackner2022approval}.} 

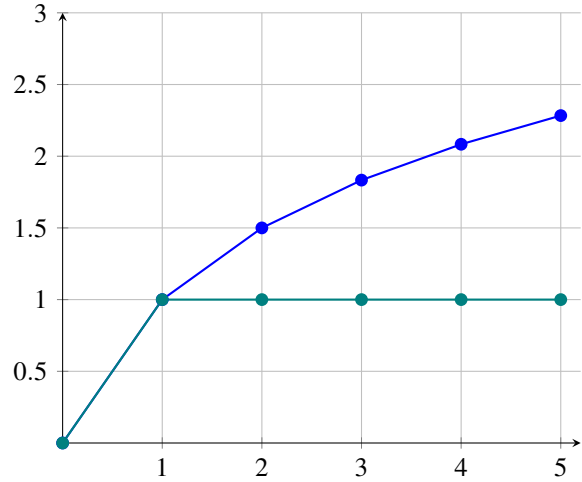
\begin{figure}
\begin{center}
\begin{tikzpicture}
  \begin{axis}[
    axis lines=middle,
    xmin=0, xmax=5.2,
    ymin=0, ymax=3,
    xtick={0,...,5},
    ytick={0,0.5,...,3},
    grid=both,
  ]
    \addplot[thick,blue,mark=*] coordinates {
      (0,0)
      (1,1)
      (2,1.5)
      (3,1.8333)
      (4,2.0833)
      (5,2.2833)
    };
    \addplot[thick,teal,mark=*] coordinates {
      (0,0)
      (1,1)
      (2,1)
      (3,1)
      (4,1)
      (5,1)
    };    
  \end{axis}
\end{tikzpicture}
\end{center}
\caption{
    Plot of the scores that the functions of equations \eqref{pav} (blue) and \eqref{cc} (green) would count for a single participant when a given number $\ell$ (here in the range $0$ to $5$) of ideas they support is selected as representative.
}
\label{fig:scores}
\end{figure}

Clearly, the nature of the solution obtained is determined by how `representativity' is conceptualized and quantified via $S$. In the literature on algorithmic deliberation support, `representativity' has been formalized in two main ways: as a form of proportionality \cite{halpern2023representation}, requiring (roughly) that when a group of participants that agrees on $m$ ideas consists of at least a share $\frac{m}{k}$ of all participants then the representation should contain at least a share $\frac{m}{k}$ of ideas supported by that group; or as a form of diversity \cite{lindeboom2025voice}, requiring that as many participants as possible have at least one idea in the computed representation that they approve of. These conceptualizations of `representativity' lead to two different ways of quantifying it, whereby the score $S(X)$ of a set of ideas $X$ is:
\begin{align}
    & \sum_{i \in N} \sum_{1 \leq \ell \leq |A_i \cap X|} \frac{1}{\ell} & \mbox{$i$ is represented by $X$ with value $1 + \frac{1}{2} + \ldots .. \frac{1}{\ell}$ (harmonic scoring)}, \label{pav} \\
    & |\{ i \in N \mid  A_i \cap X \neq \emptyset \}| & \mbox{$i$ is represented by $X$ with value $1$ or $0$ (representation's coverage)}. \label{cc}
\end{align}
See Figure \ref{fig:scores} for plots illustrating these two scoring functions.

The two papers I mentioned earlier by \citet{halpern2023representation} and \citet{lindeboom2025voice} design and study algorithms that can efficiently approximate solutions to the computational problem I defined at the beginning of the section for the two different scores of Equations \eqref{pav} and \eqref{cc}. They do so while also incorporating specific approaches to querying participants' attitudes towards ideas (see Section \ref{sec:attitudes}), in order to provide accurate estimates of the relevant representation metrics.

\subsection{Ranking ideas ideas by representativity}

The approach in the previous section assumes that only a fixed number $k$ of ideas can be displayed, and chooses the $k$ most representative ones. However, when choosing which ideas to display, one may want to consider also the {\em order} in which they are to be displayed. This gives rise to the following problem.

\begin{problem}
    {\bf {\em Representative rankings}}:
    {\em Given} a partially elicited attitude matrix, {\em find} a ranking of the set of all ideas that `fairly represents' all participants' attitudes.
\end{problem}

Also in this case, a number of approaches to the above questions are possible depending on what is understood as `representative', but a natural one involves some form of proportionality. It is clear what one would like to avoid in such a ranking: that only ideas with largest support appear at the top, thereby monopolizing the discussion. Instead, one can insist, for instance, that a group of participants whose size is a fraction $\ell$ of the total number of participants, obtains `as close as possible' to $\ell \cdot k$ of ideas in the top $k$ positions of the ranking, for any integer $k$, provided they agree on at least $\ell \cdot k$ ideas. Pursuing this intuition leads to various possible notions of {\em proportional rankings} (see \cite{liquid_feedback,skowron2017proportional} for an overview, \cite{revelrepresentative} and \cite{lederer2025proportional}). As an example, the scoring function defined in Equation \eqref{pav} above can be used to construct such a ranking.  

\medskip

Proportionality is, however, only one possible understanding of representativity in rankings and developing alternative conceptions and mathematizations for such a notion is an area of active research.

\subsection{Idea `landscapes'}

The notion of `idea landscape' (or `opinion landscape') is one of the main innovations of the platform \polis. It works based on the following intuitions about the information captured by attitude matrices (recall Figure \ref{fig:matrix}):
\begin{itemize}
    \item An attitude matrix can be viewed as an $m$-dimensional space where each issue $m$ corresponds to one dimension.
    \item If we were able to elicit a participant's attitudes on all the issues, the participant would correspond to one point in that $m$-dimensional space. This is not unlike what happens in some vote-advise software such as, e.g., \kieskompas, where participants are positioned as points on a $2$-dimensional space where spacial the proximity between two points represents ideological proximity on a number of pre-selected issues. 
\end{itemize}
One can then think of a given attitude matrix as providing information about how `close' participants are with respect to one another based on their attitudes towards the ideas proposed in the deliberation. To compute such conceptualization, three steps are currently implemented in \polis:
\begin{description}
    \item[Attitude imputation.] This occurs in cases in which the algorithms used cannot handle the fact that attitude matrices are incomplete. Instead of estimating aggregate approval scores for candidates, \polis~imputes an average attitude value to each missing value, corresponding to the issue average. This may be considered crude, and could be obviated by better elicitation or estimation methods (see Section \ref{sec:inc} below). The attitude imputation problem can also be approached to via more sophisticated prediction/inference algorithms, as done for instance in \remesh. Arguably however, imputation may be considered problematic from a democratic standpoint as it implies attributing attitudes to participants, to which they did not explicitly agree.
    
    \item[Dimensionality reduction]. The $m$-dimensional space corresponding to the attitude matrix needs to be reduced to a $2$-dimensional space for visualization on a screen. For this purpose, \polis~uses a well-known algorithm known as Principal Component Analysis (PCA). I note, however, that such algorithm is not designed for binary data, such as the ones that are collected through \polis. PCA essentially finds a subspace $\mathcal{S}$ in $d$ orthogonal dimensions, with $d < m$ (and typically $d = 2$ for visualization purposes) such that the sum of the Euclidean distances between each participant $i$ and their projection $i_\mathcal{S}$ is minimized:
    \begin{equation}
       \argmin_{\mathcal{S}} \sum_{i \in N} \|i - i_{\mathcal{S}}\|^2, \label{eq:pca} 
    \end{equation}
    where $\| \cdot \|$ 
denotes the Euclidean distance. Whether this specific optimization problem can be justified from a deliberative democratic standpoint is an interesting question. In general, in dimensionality reduction, a core issue should be to minimize some information loss (as in Equation \eqref{eq:pca}) while at the same time maintaining some fairness or representation guarantees. This would imply thinking about dimensionality reduction as a constrained optimization problem in a way analogous to the one pursued in \cite{revelrepresentative} for rankings.
   
    \item[Clustering.] To determine which participant is `close' to which other participants a clustering approach is used. \polis~uses again a well-known algorithm which essentially chooses an exogenously fixed number $k$ of `centroids' that minimizes total distance (in the underlying space) from each participant, which, remember, can be viewed as a point in the space. This corresponds to the following optimization objective:
    \begin{equation}
    \argmin_{\mathcal{C}}
        \sum_{i=1}^{k} \sum_{x \in C_i} \|x - \mathbf{\mu}_i\|^2, \label{eq:kmeans}
    \end{equation}
where $\| \cdot \|$ 
denotes the Euclidean distance, $\mathcal{C} = \{C_i \mid 1 \leq i \leq k\}$ is a partition of $N$ where each $C_i$ is a cluster and $\mu_i$ is cluster $C_i$'s centroid (i.e., mean). 

While this is a standard approach to clustering, it is not obvious that Equation \eqref{eq:kmeans} is a sound choice in a sense-making context for deliberation support.
This notion of clustering may be disadvantageous for some individuals, who may end up in a clusters whose centroid is far away from them. The literature on clustering has developed a number of alternative {\em fair clustering} algorithms based on fairer objectives than the one of Equation \eqref{eq:kmeans}, including representation objectives (see, for instance, \cite{bera2019fair,chhabra2021overview,kellerhals2024proportional}). 
\end{description}

So, at least three computational problems can underpin the sense-making approach via idea landscapes: 
\begin{problem}
    {\bf {\em Fair dimensionality reduction}}:
    {\em Given} a partially elicited attitude matrix, {\em find} a reduced 2D space on which participants can be mapped in a way that `fairly represents' their attitudes.
\end{problem}
\begin{problem}
    {\bf {\em Fair clustering}}:
    {\em Given} a partially elicited attitude matrix, {\em find} a clustering of participants that `fairly represents' their attitudes.
\end{problem}


\subsection{Handling data incompleteness} \label{sec:inc}

There are three fundamental sources of data incompleteness in deliberation support: {\em first}, participants constantly enter and exit the system, so at any given time only a subset of participants is available and can be engaged with; {\em second}, at each step in the process new ideas are contributed to the discussion, so the set of ideas actually grows over time; {\em third}, at some point the number of ideas contributed to the deliberation is too large for a single individual to be able to consider them all. 

Data incompleteness (also called `sparsity') means concretely that sense making in DP requires the use of estimates of relevant metrics rather than exact measurements of those metrics. For example, the algorithms discussed in Section \ref{sec:rep_set} above work by computing estimations of the representativity scores in Equations \eqref{cc} or \eqref{pav}. Crucially, these estimations may involve some level of adaptivity, that is, algorithmic decisions that concern, for instance, which ideas to submit to which participant, or with what probability to submit a given idea to any available participant. The challenge of sense making interacts directly with the ideas-elicitation challenge we discussed in Section \ref{sec:attitudes} and problems such as the {\em adaptive ideas-elicitation} problem. The extent to which such adaptivity is needed, and can be realized algorithmically in a democratic context is an open line of research at the interface of sequential decision-making under uncertainty \citep{sutton1998reinforcement,settles2009active} and computational social choice \citep{brandt2012computational}.\footnote{For recent contributions at this interface see \citet{halpern2023representation,lindeboom2025voice,revelrepresentative} and \citet{bachmann2024fast}.}

\subsection{Rich representations}

The three sense making approaches considered so far do not necessarily involve the processing of any linguistic information inputted by participants. They can be solely based on information collected in an attitude matrix. However, it may be argued that the texts by means of which ideas are expressed and contributed to the system contain information that can be valuable to make sense of the current state of the deliberation. This might involve, for example, reason-giving (i.e., argumentative) information which can provide insights into why certain participants hold the views expressed in the ideas they contributed. What I refer to here as `rich representations'  is any form of structured language-based representation (e.g., from simple word clouds, to complex argument maps, to generated `group statements') that can represent relevant content of the deliberation at a given time. Of special interest are representations capturing forms of reason-giving or argument. Extracting such representation from text---so-called {\em argument mining}\footnote{Here is a definition from \cite{lawrence2019argument}: ``
Argument mining is the automatic identification and extraction of the structure of inference and reasoning expressed as arguments presented in natural language.''}---is a notoriously difficult computational task, and one for which no robust solution exists to date. At the same time, novel advances in natural language processing based on large-language models (LLMs) are opening up whole new approaches to argument mining and related tasks \citep{burton2024large}, as well as to forms of automatic mediation \cite{tessler2024ai}.

\subsection{On the role of AI in DPs} 

Up until this section, the paper has refrained from using the term `artificial intelligence' (AI). This is a good place to elaborate on what I believe the role of AI in democratic deliberation support should be. Several of the algorithms I referred to, for instance for dimensionality reduction and clustering, are long-established AI algorithms that belong to the class of so-called {\em unsupervised learning} algorithms, that is, algorithms that `learn' without supervision (i.e., data labeled by humans). Furthermore, much of the academic research I cited in the paper was published in high-profile AI research venues but does not concern (with some exceptions, e.g., \cite{fish2024generative}) last-generation AI techniques that rely on methods such as large language models (LLMs) or, more generally, deep learning (DL). 

The role that DL-based techniques\footnote{In the current public and also academic discourse on AI, the terms `Deep Learning' and even `Large Language Models' are conflated with the, technically broader, term `Artificial Intelligence'. In my text I try to keep them distinct.} should play in scaling deliberation is currently actively debated, sparked by some high-profile experiments with LLM-based mediation \citep{tessler2024ai}.\footnote{See, for instance, \cite{mikhaylovskaya2024enhancing,mckinney2025five}.}
I believe that the application of DL-based techniques to scaling democratic deliberation should be pursued with a certain level of caution. The main reason is that DL-based techniques are open to the objection of being in tension with core democratic demands. The most critical tension occurs, in my view, with respect to the demand of transparency of democratic decisions, intended as the possibility of easily replicating the processes---in the case of algorithms, the computations---that lead to specific outcomes. Replicating the computation that leads to a specific output given by an LLM (for instance for moderation or what I called {\em sense making}, as in the \habermas) is unfeasible, because of the sheer number (in the trillions) of hidden parameters involved in any such computation.\footnote{This holds even when having access to the random seeds used by the LLM in the computation.}  If a computation cannot be replicated in practice, it cannot in principle be justified to a user. This has the potential to undermine the legitimacy of the democratic process that rests on such technology, at least in contexts in which high-stakes decisions are involved. 

Of course, DL-based techniques such as LLMs could find extremely useful applications in deliberation support (e.g., for automatic translation in inter-language deliberations) but a legitimacy trade-off will always be present and should be actively weighted. In general, if a core functionality is identified for a specific deliberation support software, algorithms that could deliver such functionality even approximately should perhaps be preferred to DL-based algorithms that deliver the same functionality, possibly to a higher standard, unless the legitimacy of the decision is not particularly salient as, for instance, in high-frequency, low-stakes decisions.


\section{Conclusions}

The paper has identified core functionalities---collected under what I referred to as {\em deliberation support loop}---in deliberation-support platforms (DPs), and unpacked some of the computational challenges that underpin the delivery of those functionalities in such sofware. I focused especially on the mathematical and computational dimension of the design of those functionalities and I argued that algorithms in DPs need to be designed according to democratic principles if we want them to be legitimately deployed in democratic contexts at scale. In doing so, I tried to highlight how the quest for adherence to democratic principles gives rise to novel interesting mathematical problems that need to be overcome in order for DP software to be developed responsibly. 

\medskip

I conclude with an important remark. DPs are inherently {\em socio-technical} systems, in which system behavior is determined by the interaction between algorithms with human participants. While the toolbox of computer science can build confidence in the democratic quality of algorithms used in DPs,  understanding how those algorithms will behave in practice requires insights that only empirical methods from the social sciences (sociology, psychology, social psychology, economics) can provide.\footnote{See  \cite{grossi2024enabling} for an overview of interdisciplinary challenges in digital democracy.}  




\medskip

\section*{Acknowledgments}

This research was funded by Provincie Zuid Holland, and Provincie Groningen in the context of the \href{https://polisnl.org/}{PolisNL} project. 

\medskip

Furthermore, I wish to acknowledge support from the \href{https://hybrid-intelligence-centre.nl}{Hybrid Intelligence Center}, a 10-year program funded by the Dutch
Ministry of Education, Culture and Science through the Netherlands Organisation for Scientific Research (NWO) and by the European Union under the Horizon Europe project \href{https://perycles-project.eu/}{Perycles} (Participatory Democracy that Scales). Views and opinions expressed are however those of the author only and do not necessarily reflect those of the European Union or the European Research Executive Agency (REA). Neither the European Union nor the granting authority can be held responsible for them.

\smallskip
\begin{center}
\includegraphics[width=0.5\textwidth]{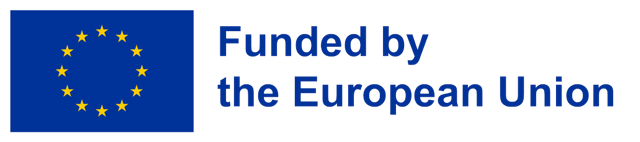}
\end{center}

I wish to thank Jan Behrens and Andreas Nitsche for useful feedback on an earlier version of this manuscript.



\begin{thebibliography}{}

\bibitem[Bachmann et~al., 2024]{bachmann2024fast}
Bachmann, F., Sarasua, C., and Bernstein, A. (2024).
\newblock Fast and adaptive questionnaires for voting advice applications.
\newblock In {\em Joint European Conference on machine learning and knowledge
  discovery in databases}, pages 365--380. Springer.

\bibitem[Behrens et~al., 2014]{liquid_feedback}
Behrens, J., Kistner, A., Nitsche, A., and Swierczek, B. (2014).
\newblock {\em Principles of Liquid Feedback}.
\newblock Interaktieve Demokratie.

\bibitem[Bera et~al., 2019]{bera2019fair}
Bera, S., Chakrabarty, D., Flores, N., and Negahbani, M. (2019).
\newblock Fair algorithms for clustering.
\newblock {\em Advances in Neural Information Processing Systems}, 32.

\bibitem[Brandt et~al., 2012]{brandt2012computational}
Brandt, F., Conitzer, V., and Endriss, U. (2012).
\newblock Computational social choice.
\newblock {\em Multiagent systems}, 2:213--284.

\bibitem[Burton et~al., 2024]{burton2024large}
Burton, J.~W., Lopez-Lopez, E., Hechtlinger, S., Rahwan, Z., Aeschbach, S.,
  Bakker, M.~A., Becker, J.~A., Berditchevskaia, A., Berger, J., Brinkmann, L.,
  et~al. (2024).
\newblock How large language models can reshape collective intelligence.
\newblock {\em Nature human behaviour}, 8(9):1643--1655.

\bibitem[Chhabra et~al., 2021]{chhabra2021overview}
Chhabra, A., Masalkovait{\.e}, K., and Mohapatra, P. (2021).
\newblock An overview of fairness in clustering.
\newblock {\em IEEE Access}, 9:130698--130720.

\bibitem[Dahl, 2008]{dahl2008polyarchy}
Dahl, R.~A. (2008).
\newblock {\em Polyarchy: Participation and opposition}.
\newblock Yale university press.

\bibitem[Denning et~al., 1989]{denning1989computing}
Denning, P.~J., Comer, D.~E., Gries, D., Mulder, M.~C., Tucker, A., Turner,
  A.~J., and Young, P.~R. (1989).
\newblock Computing as a discipline.
\newblock {\em Computer}, 22(2):63--70.

\bibitem[Fish et~al., 2024]{fish2024generative}
Fish, S., G{\"o}lz, P., Parkes, D.~C., Procaccia, A.~D., Rusak, G., Shapira,
  I., and W{\"u}thrich, M. (2024).
\newblock Generative social choice.
\newblock In {\em Proceedings of the 25th ACM Conference on Economics and
  Computation}, pages 985--985.

\bibitem[Grossi et~al., 2024]{grossi2024enabling}
Grossi, D., Hahn, U., M{\"a}s, M., Nitsche, A., Behrens, J., Boehmer, N.,
  Brill, M., Endriss, U., Grandi, U., Haret, A., et~al. (2024).
\newblock Enabling the digital democratic revival: A research program for
  digital democracy.
\newblock {\em arXiv preprint arXiv:2401.16863}.

\bibitem[Halpern et~al., 2023]{halpern2023representation}
Halpern, D., Kehne, G., Procaccia, A.~D., Tucker-Foltz, J., and W{\"u}thrich,
  M. (2023).
\newblock Representation with incomplete votes.
\newblock In {\em Proceedings of the AAAI Conference on Artificial
  Intelligence}, volume~37, pages 5657--5664.

\bibitem[Kellerhals and Peters, 2024]{kellerhals2024proportional}
Kellerhals, L. and Peters, J. (2024).
\newblock Proportional fairness in clustering: A social choice perspective.
\newblock {\em Advances in Neural Information Processing Systems},
  37:111299--111317.

\bibitem[Lackner and Skowron, 2022]{lackner2022approval}
Lackner, M. and Skowron, P. (2022).
\newblock Approval-based committee voting.
\newblock In {\em Multi-Winner Voting with Approval Preferences}, pages 1--7.
  Springer.

\bibitem[Lawrence and Reed, 2019]{lawrence2019argument}
Lawrence, J. and Reed, C. (2019).
\newblock Argument mining: A survey.
\newblock {\em Computational linguistics}, 45(4):765--818.

\bibitem[Lederer, 2025]{lederer2025proportional}
Lederer, P. (2025).
\newblock Proportional representation in rank aggregation.
\newblock {\em arXiv preprint arXiv:2508.16177}.

\bibitem[Lindeboom et~al., 2025]{lindeboom2025voice}
Lindeboom, F., Brehm, M., Grossi, D., and Murukannaiah, P. (2025).
\newblock A voice for minorities: diversity in approval-based committee
  elections under incomplete or inaccurate information.
\newblock {\em arXiv preprint arXiv:2506.10843}.

\bibitem[McKinney and Chwalisz, 2025]{mckinney2025five}
McKinney, S. and Chwalisz, C. (2025).
\newblock Five dimensions of scaling democratic deliberation: With and beyond
  ai.
\newblock {\em Brussels: DemocracyNext. Retrieved April}, 19:2025.

\bibitem[Mikhaylovskaya, 2024]{mikhaylovskaya2024enhancing}
Mikhaylovskaya, A. (2024).
\newblock Enhancing deliberation with digital democratic innovations.
\newblock {\em Philosophy \& Technology}, 37(1):3.

\bibitem[Narayanan, 2023]{narayanan2023understanding}
Narayanan, A. (2023).
\newblock Understanding social media recommendation algorithms.

\bibitem[Rawls, 2017]{rawls2017theory}
Rawls, J. (2017).
\newblock A theory of justice.
\newblock In {\em Applied ethics}, pages 21--29. Routledge.

\bibitem[Revel et~al., 2025]{revelrepresentative}
Revel, M., Milli, S., Lu, T., Watson-Daniels, J., and Nickel, M. (2025).
\newblock Representative ranking for deliberation in the public sphere.
\newblock In {\em Forty-second International Conference on Machine Learning}.

\bibitem[Robert~III et~al., 2020]{robert2020robert}
Robert~III, H.~M., Honemann, D.~H., Balch, T.~J., Seabold, D.~E., and Gerber,
  S. (2020).
\newblock {\em Robert's rules of order newly revised}.
\newblock PublicAffairs.

\bibitem[Rosa et~al., 2025]{rosa2025harnessing}
Rosa, P., Guimares, P.~A., et~al. (2025).
\newblock Harnessing innovation on online deliberation.
\newblock {\em European Commission, Joint Research Center}.

\bibitem[Salganik and Levy, 2015]{salganik2015wiki}
Salganik, M.~J. and Levy, K.~E. (2015).
\newblock Wiki surveys: Open and quantifiable social data collection.
\newblock {\em PloS one}, 10(5):e0123483.

\bibitem[Settles, 2009]{settles2009active}
Settles, B. (2009).
\newblock Active learning literature survey.

\bibitem[Skowron et~al., 2017]{skowron2017proportional}
Skowron, P., Lackner, M., Brill, M., Peters, D., and Elkind, E. (2017).
\newblock Proportional rankings.
\newblock In {\em Proceedings of the 26th International Joint Conference on
  Artificial Intelligence}, pages 409--415.

\bibitem[Small et~al., 2021]{small2021polis}
Small, C., Bjorkegren, M., Erkkil{\"a}, T., Shaw, L., and Megill, C. (2021).
\newblock Polis: Scaling deliberation by mapping high dimensional opinion
  spaces.
\newblock {\em Recerca: revista de pensament i an{\`a}lisi}, 26(2).

\bibitem[Sutton et~al., 1998]{sutton1998reinforcement}
Sutton, R.~S., Barto, A.~G., et~al. (1998).
\newblock {\em Reinforcement learning: An introduction}, volume~1.
\newblock MIT press Cambridge.

\bibitem[Tessler et~al., 2024]{tessler2024ai}
Tessler, M.~H., Bakker, M.~A., Jarrett, D., Sheahan, H., Chadwick, M.~J.,
  Koster, R., Evans, G., Campbell-Gillingham, L., Collins, T., Parkes, D.~C.,
  et~al. (2024).
\newblock Ai can help humans find common ground in democratic deliberation.
\newblock {\em Science}, 386(6719):eadq2852.

\end{thebibliography}

\end{document}